\documentclass[conference]{IEEEtran}
\IEEEoverridecommandlockouts
\usepackage{blindtext, graphicx}
\usepackage{caption}
\ifCLASSINFOpdf
\else
\fi
\usepackage[cmex10]{amsmath}

\usepackage{float}
\begin{document}
%
% paper title
% can use linebreaks \\ within to get better formatting as desired
\title{Agent Inspired Trading Using Recurrent Reinforcement Learning and LSTM Neural Networks }

% author names and affiliations
% use a multiple column layout for up to three different
% affiliations
\author{\IEEEauthorblockN{David W. Lu}\thanks{At the time of completion of this article, the author works for Bank of America Merrill Lynch. The views and opinions expressed in this article are those of the author and do not necessarily reflect the views or position of Bank of America Merrill Lynch}
%\IEEEauthorblockA{School of Electrical and\\Computer Engineering\\
%Georgia Institute of Technology\\
%Atlanta, Georgia 30332--0250\\
Email: davie.w.lu@gmail.com}

%\and
%\IEEEauthorblockN{Homer Simpson}
%\IEEEauthorblockA{Twentieth Century Fox\\
%Springfield, USA\\
%Email: homer@thesimpsons.com}
%\and
%\IEEEauthorblockN{James Kirk\\ and Montgomery Scott}
%\IEEEauthorblockA{Starfleet Academy\\
%San Francisco, California 96678-2391\\
%Telephone: (800) 555--1212\\
%Fax: (888) 555--1212}}

% conference papers do not typically use \thanks and this command
% is locked out in conference mode. If really needed, such as for
% the acknowledgment of grants, issue a \IEEEoverridecommandlockouts
% after \documentclass

% for over three affiliations, or if they all won't fit within the width
% of the page, use this alternative format:
% 
%\author{\IEEEauthorblockN{Michael Shell\IEEEauthorrefmark{1},
%Homer Simpson\IEEEauthorrefmark{2},
%James Kirk\IEEEauthorrefmark{3}, 
%Montgomery Scott\IEEEauthorrefmark{3} and
%Eldon Tyrell\IEEEauthorrefmark{4}}
%\IEEEauthorblockA{\IEEEauthorrefmark{1}School of Electrical and Computer Engineering\\
%Georgia Institute of Technology,
%Atlanta, Georgia 30332--0250\\ Email: see http://www.michaelshell.org/contact.html}
%\IEEEauthorblockA{\IEEEauthorrefmark{2}Twentieth Century Fox, Springfield, USA\\
%Email: homer@thesimpsons.com}
%\IEEEauthorblockA{\IEEEauthorrefmark{3}Starfleet Academy, San Francisco, California 96678-2391\\
%Telephone: (800) 555--1212, Fax: (888) 555--1212}
%\IEEEauthorblockA{\IEEEauthorrefmark{4}Tyrell Inc., 123 Replicant Street, Los Angeles, California 90210--4321}}

% use for special paper notices
%\IEEEspecialpapernotice{(Invited Paper)}

% make the title area
\maketitle

\begin{abstract}
%\boldmath
With the breakthrough of computational power and deep neural networks, many areas that we haven't explore with various techniques that was researched rigorously in past is feasible. In this paper, we will walk through possible concepts to achieve robo-like trading or advising. In order to accomplish similar level of performance and generality, like a human trader, our agents learn for themselves to create successful strategies that lead to the human-level long-term rewards. The learning model is implemented in Long Short Term Memory (LSTM) recurrent structures with Reinforcement Learning or Evolution Strategies acting as agents The robustness and feasibility of the system is verified on GBPUSD trading.

\end{abstract}
% IEEEtran.cls defaults to using nonbold math in the Abstract.
% This preserves the distinction between vectors and scalars. However,
% if the journal you are submitting to favors bold math in the abstract,
% then you can use LaTeX's standard command \boldmath at the very start
% of the abstract to achieve this. Many IEEE journals frown on math
% in the abstract anyway.

% Note that keywords are not normally used for peerreview papers.

% For peer review papers, you can put extra information on the cover
% page as needed:
% \ifCLASSOPTIONpeerreview
% \begin{center} \bfseries EDICS Category: 3-BBND \end{center}
% \fi
%
% For peerreview papers, this IEEEtran command inserts a page break and
% creates the second title. It will be ignored for other modes.
\IEEEpeerreviewmaketitle

\section{Introduction}
Many of the machine learning or artificial intelligence techniques can be trace back as early as the 1950s. Evolving from the study of pattern recognition and computational learning theory, researchers explore and study the construction of algorithms that can learn and predict on data. With the predictions,  researchers came across the idea of of a learning system that can decide something, that adapts its behavior in order to maximize a signal from its environment. This was the creation of a "hedonistic" learning system.[1] The idea of this learning system may be viewed as Adaptive Optimal Control, nowadays we call it reinforcement learning [2]. In order to accomplish similar level of performance and generality, like a human, we need to construct and learn the knowledge directly from raw inputs, such as vision, without any hand-engineered features, this can be achieved by deep learning of neural networks. Combining the two, some simply refer it to deep reinforcement learning, which could create an artificial agent that is as close as we can to sanely call it true "artificial intelligence".

In this paper, we'll focus with direct reinforcement or recurrent reinforcement learning to refer to algorithms that do not have to learn a value function in order to derive a policy. Some researchers called policy gradient algorithms in a Markov Decision Process framework as direct reinforcement to generally refer to any reinforcement learning algorithm that does not require learning a value function. Herein, we will focus on recurrent reinforcement learning. Methods such as dynamic programming[3], TD Learning[4] or Q-Learning[5] had been the focus of most modern researches. These method attempt to learn a value function. Actor-Critic methods[6], which is an intermediate between direct reinforcement and value function methods, in that the "critic" learns a value function which is then used to update the parameters of the "actor". 

Why we chose to focus on recurrent reinforcement learning? Though much theoretical progress has been made in the recent years, there had been few public known applications in the financial world. We as start-ups, quantitative hedge funds, client driven investment services, wealth management companies, and most recently robo-advisers, had been focusing on financial decision making problems to trade on its own. Within the reinforcement learning community, much attention is actually given to the question of learning policies versus learning value functions. The value function approach as described earlier had dominated the field throughout the last thirty years. The approach had worked well in many applications, alpha Go, training a Helicopter to name a few. However, the value function approach suffers from several limitations. Q-Learning is in the context of action spaces and discrete state. In many situations, this will suffer the "curse of dimensionality" When Q-Learning is extended to function approximators, researchers had shown it fail to converge using a simple Markov Decision Process. Brittleness which means small change in the value function can produce large changes in the policy. In the trading signal world, the data can be in presence of large amounts of noise and nonstationarity in the datasets, which could cause severe problems for a value function approach.

Recurrent reinforcement learning can provide immediate feedback to optimize the strategy, ability to produce real valued actions or weights naturally without resorting to the discretization necessary for value function approach. There are other portfolio optimization techniques such as evolution strategies and linear matrix inequilities which relies on predicting the covariance matrix and optimize.  For all optimization problems or in reinforcement learning set up, we need an objective, and such objective can be formulated in terms of risk or rewards. Moody et al.[7] shown that how differential forms of the Sharpe Ratio and Downside Deviation Ratio can be formulated to enable efficient on-line learning with recurrent reinforcement learning, Lu[8] had shown using Linear Matrix Inequilities can beat the risk free rate, and Deng et al.[9] had shown max return can be used as objective in recurrent reinforcement learning as well as using deep learning transformations to initialize the features. 

To extend the recurrent structure, we'll further discuss in this paper how the back propagation through time method is exploited to unfold the recurrent neural network as a series of time-dependent stacks without feedback. As discuss in [9], gradient vanishing issue is inevitably true in these structures. This was because the unfolded neural networks exhibits extremely deep structures on the feature learning and also the time expansion parts. We introduce the Long Short Term Memory (LSTM) to handle this deficiency. We will discuss the characteristics of LSTM as well as thoughts and techniques such as Dropouts [10] to test.  This strategy provides a chance to forecast the final objective and improve the learning efficiency.

The recurrent reinforcement learner requires to optimize the objective through gradient ascent. In this paper, we will also explore literature in Evolution Strategies [11] and Nelder-Mead method [12] as to search for the gradient or so called direct search or derivative-free methods.

Finally, the Trading Systems will be tested among S\&P 500, EURUSD, and commodity futures market. The remaining parts of this part are organized as follows. Section II, we will walk through how we construct the trading agents, Section III reveals how we construct the recurrent layers in plain recurrent and LSTM. Additionally, how dropouts can affect the training and reduce gradient vanishing issues. Section IV, we will talk about gradient ascent, evolution strategies and Helder-Mead Methods. Section V, we will detail the test results and comparison of methods listed in Section II to IV. Section VI concludes his paper and provides thoughts on future directions.

%\footnote{At the time of writing this article, the author works for JP Morgan Chase \& Co. The views and opinions expressed in this article are those of the authors and do not necessarily reflect the views or position of JP Morgan Chase \& Co}
%\subsection{Subsection Heading Here}
%\blindtext

\section{Recurrent Reinforcement Learning}
To demonstrate the feasibility of agents that trades, we consider agents that trade fixed position size on a single security. The methods described here can be generalized to more sophisticated agents that trades or optimize a portfolio, trades varying quantities of a security, allocate assets continuously or manage multiple asset portfolios. We'll discuss this further in separate sessions. See [13] for some initial discussions. 

Intuitively, we find a objective function so that the agent knows what we are trying to maximize or minimize. Most modern fund managers attempt to maximize risk-adjusted return using the Sharpe Ratio, as suggested by modern portfolio theory. The Sharpe Ratio is define as follows[14]:
\begin{equation}
S_T = \frac{Average(R_t)}{Standard Deviation (R_t)} = \frac{E[R_t]}{\sqrt{E[{R_t}^2]-(E[R_t])^2}}
\end{equation}
where \begin{math}R_t\end{math} is the return on investment for trading period t and E[.] denotes the expectation. In modern portfolio theory, higher Sharpe Ratio rewards investment strategies that rely on less volatile trends to make a profit. As discussed earlier, there are other functions or ratios we can use, however for ease of demonstration purposes we will use Sharpe Ratio and Downside Deviation Ratio in this article.

Next step we need to define how a agent would trade. The trader would take a long, neutral or a short position. A long position is entering a purchase of some quantity of a security, while a short position is triggered by selling the security. Herein, we will follow mostly the notation in [7][15] for the ease of explaining and reconciliation. Let's define \begin{math}F_t\in [-1, 0. 1]\end{math} represents the trading positions at time t. A long position when \begin{math}F_t>0\end{math}. In this case, the trader buys the security at price \begin{math}P_t\end{math}, and hopes that the prices goes up on period t+1. A short position is when \begin{math}F_t<0\end{math}. In this case, the trader short sale (borrow to sell) the security at price \begin{math}P_t\end{math}, and hopes that the prices goes down on period t+1 so that trader can buy it back to return the security that it borrowed. Intuitively, one can use a Tanh function to represent this set up since it's goes from -1 to 1. 

We define the trader function in a simple form of:
\begin{equation}
F_t = tanh(w^Tx_t)
\end{equation}
where \begin{math}x_t = [r_{t-m+1},... r_t]\end{math} and the return \begin{math}r_t = p_t-p_{t-1}\end{math}
Note that the trader function can also add in a bias term b and the last trading decision with parameter u to add into the regression. The latest trading decision with parameter can discourage the agent to frequently change the trading positions and to avoid huge transaction costs. We can then rewrote the equation to 
\begin{equation}
F_t = tanh(w^Tx_t+b+uF_{t-1})
\end{equation}

Adding the number of shares as s with transaction cost c, we can then write the return at time t as 
\begin{equation}
R_t = s(F_{t-1}r_t-c|F_t-F_{t-1}|)
\end{equation}

With the above elements set up, we can now try to maximize the Sharpe Ratio using Gradient Ascent or other methods which we'll discuss further in Section IV to the find the optimal weights for the agent to use. Again let's think through given trading system model \begin{math}F_t\end{math}, the goal is to adjust the parameter or weights w in order to maximize \begin{math}S_T\end{math}. We can write the weights as follows:
\begin{equation}
w_t = w_{t-1}+\rho\frac{dS_t}{dw_t}=w_{t-1}+\Delta w
\end{equation}
where \begin{math}w_t\end{math} is any weight of the network at time t, \begin{math}S_t\end{math}is the measure we'd like to maximize or minimize, and \begin{math}\rho\end{math} is an adjustable learning rate.

Examining the derivatives of \begin{math}S_T\end{math} or gradient with respect to the weight w over a sequence of T periods is:
\begin{equation}
\frac{dS_T}{dw} = \sum_{t=1}^{T}\frac{dS_T}{dR_t}\bigg\{\frac{dR_t}{dF_t}\frac{dF_t}{dw}+\frac{dR_t}{dF_{t-1}}\frac{dF_{t-1}}{dw}\bigg\}
\end{equation}
The trader can then be optimized in batch mode by repeatingly compute the value of $S_T$ on forward passes through the data with:
\begin{equation}
\frac{dR_t}{dF_t} = -scsign(F_t=F_{t-1})
\end{equation}
\begin{equation}
\frac{dR_t}{dF_{t-1}}=r_t+scsign(F_t=F_{t-1})
\end{equation}
\begin{equation}
\frac{dF_t}{dw}= \frac{\partial F_t}{\partial w}+\frac{\partial F_t}{\partial F_{t-1}}\frac{dF_{t-1}}{dw}
\end{equation}
Due to the inherent recurrence, the quantitites $dF_t/dw$ are total derivatives that depend upon the entire sequence of previous time periods. In other words, $dF_t/dw$ is recurrent and depends on all previous values. Though it does slow down the gradient ascent but due to modern computational power and range of samples, it does not present insuperable burden. To correctly compute and optimize these total derivatives, we can deploy a similar bootstrap method as in Back-Propagation Through Time(BPTT)[16]. Alternatively, one can use a simple on-line stochastic optimization by consider only the term in (6) that depends on the most recent realized return $R_t$ during a forward pass through the data. The equation in (6) becomes:
\begin{equation}
\frac{dS_t}{dw} \approx \sum_{t=1}^{T}\frac{dS_t}{dR_t}\bigg\{\frac{dR_t}{dF_t}\frac{dF_t}{dw}+\frac{dR_t}{dF_{t-1}}\frac{dF_{t-1}}{dw}\bigg\}
\end{equation}
Such an algorithm performs a stochastic optimization or effectively making the algorithm a stochastic gradient ascent. As we previously mentioned, there are other methods to maximize the objective function. We'll discuss that further in section IV.

We also tested the weight decay variant of the gradient ascent learning algorithm as described in [15] to verify the performance. Using the weight decay, (5) becomes:
\begin{equation}
w_t = w_{t-1}+\rho\frac{dS_t}{dw_t}-\nu w_{t=-1}=w_{t-1}(1-\nu)+\Delta w
\end{equation}
where \begin{math}\nu\end{math} is the co-efficient of weight decay. Adding the weight decay improves neutral network performance based on the fact that smaller weights will have less tendency to over-fit the noise in the data. Similar to the findings in [15], the weight decay will not help single layer neural networks since it's theoretically for the purpose to simplify the rule learned by the neural network and prevent the neural network from memorizing noise in the data. The next section will introduce the deep learning transformation and dropouts to better fine tune the performance.

Although Sharpe ratio is the most widely used risk-adjusted metric, it provides rankings that it is counter-intuitive investors' sense of risk because the use of variance or $R^2_t$ as risk measure does not distinguish between upside and downside risk, therefore penalize both large positive or negative returns. To most investors, the risk refers to returns in a portfolio that decreases its profitability.
In this paper, we will experiment both signals with recurrent neural network and downside deviation ratio to protect downside risk. 

Similar to equation (1), we can define downside deviation ratio as follows:
\begin{equation}
D_T = \frac{Average(R_t)}{DD_T} = \frac{E[R_t]}{\sqrt{E[min[R_t,0]^2]}}
\end{equation}
Equation (5) becomes
\begin{equation}
w_t = w_{t-1}+\rho\frac{dD_t}{dw_t}=w_{t-1}+\Delta w
\end{equation}
Computationally, it will be easier if 0 here is described as a very small number. We will check the performance of downside deviation ratio and Sharpe ratio in Section V.
%\subsection{Subsection Heading Here}

\section{LSTM For Informative Feature Learning}
To further this research, we attempt to find efficient algorithms that takes the decision objective into account when estimating either the covariance matrix [17] or the features [9]. As an example of the former, the Directed Principal Component Analysis [17] is stated for estimating the covariance matrix with the decision objective in mind. Such method is useful for portfolio estimations and predictions. The latter, which we attempt to use a deep neural network transformation or a fuzzy learning method to help understand the signals we feed into the recurrent reinforcement learning structure[9]. Herein, we will explore using Long Short Term Memory.

We implemented LSTM(Long Short Term Memory) [18] to understand and dynamically sense the market condition and use it for informative feature learning. In theory, the appeals of Recurrent Neural Networks is the idea that they might be able to connect previous information to the present task we are aiming to achieve. Unfortunately, in practice, it is possible for the gap between the relevant information and point where it is needed become very large. As the gap grows, RNNs become unable to learn to connect the information [19]. LSTM was first introduced in 1997 [18] to resolve the difficulties to model long sequences. The fundamental issue was that gradients propagated over many stages tend to either vanish or explode. In a traditional recurrent neural network, during the gradient back-propagation phase, the gradient signal can end up being multiplied by humongous times perhaps as many times as of the timesteps by the weight matrix associated with the connections between the neurons of the recurrent hidden layer. In other words, the magnitude of weights in the transition matrix can have a large impact on the learning process. If the weights in this matrix are small it leads to vanishing gradients where the gradient signal gets so small that learning either becomes very slow or stops working altogether. In defiance of this, if the weights in this matrix are large where the gradient signal is large, where we often refer this as exploding gradients.

Previously, we talked about the issues based on a recurrent neural network. These issues are the main motivation behind the LSTM model which introduces a new structure called a memory cell. A memory cell is built with four main elements: an input gate, a neuron with a self-recurrent connection, a forget gate and an output gate. The self-recurrent connection ensure the state of a memory cell can remain constant from one timestep to another. The input gate allows incoming signal to alter the state of the memory cell or block it. The output gate can allow the state of the memory cell to have an effect on other neurons or prevent it. Finally, the forget gate can modulate the memory cell’s self-recurrent connection, allowing the cell to remember or forget its previous state, as needed.
We may wonder why does a LSTM have a forget gate when their purpose is to link distant occurrences to a final output. When we are analyzing a time series and comes to the end of it, for example, you may have no reason to believe that the next time instance has any relationship to the previous, and therefore the memory cell should be set to zero before the next instance. In Figure 1, we can see how gates work, with straight lines representing closed gates and open circles representing open gates. The lines and circles running horizontal up the hidden layer are the forget gates.

% needed in second column of first page if using \IEEEpubid
%\IEEEpubidadjcol

% An example of a floating figure using the graphicx package.
% Note that \label must occur AFTER (or within) \caption.
% For figures, \caption should occur after the \includegraphics.
% Note that IEEEtran v1.7 and later has special internal code that
% is designed to preserve the operation of \label within \caption
% even when the captionsoff option is in effect. However, because
% of issues like this, it may be the safest practice to put all your
% \label just after \caption rather than within \caption{}.
%
% Reminder: the "draftcls" or "draftclsnofoot", not "draft", class
% option should be used if it is desired that the figures are to be
% displayed while in draft mode.
%
\begin{figure}[h]
%\DeclareGraphicsExtensions{.png,.pdf}
\centering
\includegraphics[width=2.5in]{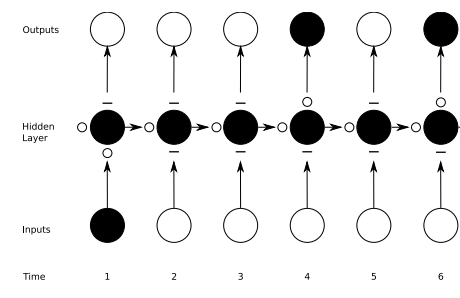}
% where an .eps filename suffix will be assumed under latex, 
% and a .pdf suffix will be assumed for pdflatex; or what has been declared
% via \DeclareGraphicsExtensions.
\captionsetup{justification=centering}
\caption{Example of a LSTM Recurrent Neural Network}
%\label{fig_sim}

\end{figure}

With the concept in mind, let's walk through the mathematical expressions. We'll try to use the notations as simple as possible here to explain. Please see [20] for further readings and detailed discussions. Note that the notation used in this section is not the same representation if any overlap as in Section II. Here $x^t$ is the input vector at time t, $h^t$ is the hidden layer vector, the W are input weight matrices and U are the recurrent weight matrices and b are bias vectors. Functions $\sigma$, m, and n are point-wise non-linear activation functions. 
Logistic sigmoid $(\frac{1}{1+e^{-x}})$ is used for activation functions of the gates or $\sigma$ and hyperbolic tangent $tanh$ is used as the block input and out activation functions (m, n). Finally, the point-wise multiplication of two vectors is denoted with $\odot$ We can write the expressions as follows:
\begin{equation}
block\:input:\quad y^t = m(W_yx^t+U_yh^{t-1}+b_y)
\end{equation}
\begin{equation}
input\:gate:\quad i^t = \sigma(W_ix^t+U_ih^{t-1}+b_i)
\end{equation}
\begin{equation}
forget\:gate:\quad f^t = \sigma(W_fx^t+U_fh^{t-1}+b_f)
\end{equation}
\begin{equation}
cell\:internal\:state:\quad c^t = i^t\odot y^t+f^t\odot c^{t-1}
\end{equation}
\begin{equation}
output\:gate:\quad o^t = \sigma(W_ox^t+U_oh^{t-1}+b_o)
\end{equation}
\begin{equation}
block\:output:\quad z^t = o^t\odot n(c^t)
\end{equation}
There are other variant of LSTM one can explore such as LSTM with peepholes [21] and more recently Associative LSTM [22]. We chose LSTM with forget gates as it is a simple yet commonly used LSTM configuration and fits our purpose.

LSTM recurrent neural networks contain various non-linear hidden layers and this makes them very expressive and can learn complicated relationships between their inputs and outputs. However, the complicated relationships will be result of sampling noise, so they will exists in the training set but not in the real data even if it's drawn from the same distribution. As explained in [23], this leads to overfitting. One of the regularization techniques called Dropout is to address this issue. It will prevent overfitting and provides a way to combine many different neural network architectures efficiently. Interestingly, it turns out that Dropouts will not work well with RNNs and LSTMs unless apply it correctly. [24] showed us how to correctly applied dropouts to LSTM. The main idea is to apply the dropout operator only to the non-recurrent layer. Figure 2. shows the dropouts only applies to the dash arrows but not the solid arrows.

\begin{figure}[h]
%\DeclareGraphicsExtensions{.png,.pdf}
\centering
\includegraphics[width=2.5in]{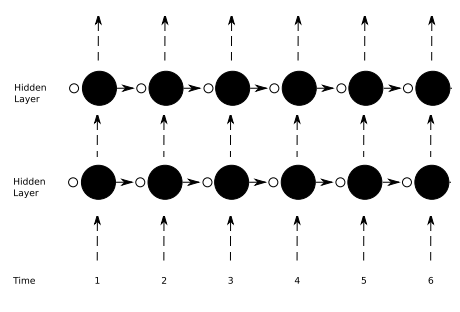}
% where an .eps filename suffix will be assumed under latex, 
% and a .pdf suffix will be assumed for pdflatex; or what has been declared
% via \DeclareGraphicsExtensions.
\captionsetup{justification=centering}
\caption{Example of applying Dropouts on LSTM}
%\label{fig_sim}

\end{figure}

Mathematically, let's assume the equation in (17) has total L Layers in our deep LSTM. The block output becomes $z^t_L$. $z^t_l$ is a hidden state in layer l in timestep t.The deterministic state transition is a function RNN: $z^t_{l-1},z^{t-1}_l\rightarrow z^t_l$ where D is the dropout operation that sets a random subset of its argument to zero, D($z^t_{l-1}$). An input of a dropout is normally a float between 0 and 1. where we define the percentage of the units to drop for the linear transformation of the inputs.

To utilize the benefits of recurrent neural networks, we incorporated LSTM and dropout to tackle the challenges in a recurrent neural network. This is our step 2 to create artificial agents that can achieve a similar level of performance and generality. In this section, we created feature learning framework to construct informative feature representation for the agent to consume. In the next section, we will swim back to the reinforcement learning to fathom methodologies on solving for the weights based on our objectives.

\section{Review of Gradient Ascent, Evolution Strategies, and Other Gradient Search}
In the finance world, being able to explain and interpret intuitively on the assumptions and models we deployed are equally important as the effectiveness and complexity of the model itself. We discussed Gradient Ascent thoroughly in finding the optimal Sharpe Ratio in Section II. In this section, we will explore other methodologies which are more closer to "Black-box" Optimization. The motivation behind these type of methods is that many real world optimization problems are too complex to model directly. Numerous algorithms in this area have been developed and applied in the past thirty years and in many cases providing near optimal solutions to extreme complicated tasks. Heuristic methods, which is any approach to problem solving, learning, or discovery that employs a practical method not guaranteed to be optimal or perfect, but sufficient for the immediate goals. When finding an optimal solution is impossible or impractical, heuristic methods can be used to speed up the process of finding a satisfactory solution. At times, the near-infeasibility of achieving globally optimal solutions requires a fair amount of heuristics in black-box algorithms, therefore, this may results in sometimes immediately seeing high-performance with methods that perhaps intuitively hard to explain. Though in this paper, we presented only agent based trading limited to a single asset, we studied these methods to prepare for later extension in portfolio optimization. 

The problem of black-box optimization had evolved to a wide variety of approaches. A first class of method is Nelder-Mead [25] which was presented back in 1965, this was inspired by classic optimization methods, such as simplex methods. More Heuristic methods as previously mentioned are inspired by natural evolution. These are a class of stochastic optimization and adaptation techniques and had been developed from the early 1950s on. Including a broad class of genetic algorithms  diﬀerential evolution, estimation of distribution algorithms, particle swarm optimization [26], the cross-entropy method [27], covariance matrix adaptation evolution strategy CMA-ES [28], which is considered by many to be the "industry standard" of evolutionary computation, Natural Evolution Strategies, and Evolino[29]. 

In optimization, a problem is typically speciﬁed by a set of n parameters $x_1,...x_n$ and an objective function $f$, which is also called a ﬁtness function in the context of evolutionary algorithms. The goal of the optimization process is to ﬁnd a set of n variables $w_1,...w_n$ such that the objective function is optimized. Without loss of generality, it is sufﬁcient to consider only minimization tasks, since maximizing $f$ is equivalent to minimizing $-f$. This will be useful in actual implementation in various optimization libraries in Python, Matlab or R. 

Each of the evolutionary algorithm we mentioned above is designed with a different methodology. Despite their differences, all evolutionary algorithms are heuristic population-based search procedures that incorporate random variation and selection. In each iteration i, also called a generation, an evolutionary algorithm typically generates $\lambda$ offspring from $\mu$ parents. Every offspring is generated by copying a parent and adding a mutation vector z to the parent’s parameters x. In evolutionary programming and evolution strategies, all components of z are typically Gaussian-distributed with mean zero and standard deviation. The standard deviation is also called the step size. By evaluating the objective function, evolutionary algorithms assign a ﬁtness value to each offspring and select offspring as parents for the next iteration(generation).A very important feature of evolutionary programming and evolution strategies is their ability to self-adapt the step size.It is very common for both evolutionary programming and evolution strategies to perform self-adaptation by considering standard deviation at iteration i as an additional parameter.

The Evolution Strategies, culminating with CMA-ES, were design to cope with high-dimensional continuous-valued domains. The algorithm framework has been developed extensively over the years to include self-adaptation of the search parameters, and the representation of correlated mutations by the use of a full covariance matrix. This allowed the framework to capture interrelated dependencies by exploiting the covariances while ‘mutating’ individuals for the next generation. While evolution strategies prove to be an eﬀective framework for black-box optimization, their ad hoc procedures remain heuristic in nature. Many literature had prove that to thoroughly analyzing the actual dynamics of the procedure turns out to be difficult[11][30]

Natural Evolution Strategies (NES)[31] are evolution-strategy inspired blackbox optimization algorithms, which instead of maintaining a population of search points, it is iteratively updating a search distribution. These type of methodology used natural gradient to update search distribution in the direction of higher expected fitness. Like CMA-ES, they can be cast into the framework of evolution strategies.  Natural evolution strategies has been successfully applied to black box optimization [28], as well as for the training of the recurrent weights in recurrent neural networks [29]. 

We are particularly interested in training Recurrent Neural Networks with Evolino. Evolino calculates weights to the nonlinear, and hidden nodes while computing optimal linear mapping from hidden state to output. We will compare against linear matrix inequilities [8] [32] in optimization of portfolio optimization in future works. 

We quickly reviewed the optimization methods, it is optimal for us initially to use Gradient Ascent as described in section I. It is easier for us to explain and implement as an agent based reinforcement learning model. Nevertheless, it is feasible to implement using evolution strategies for more complex and multi-asset optimization.

\section{EMPIRICAL RESULTS}
This section displays our empirical results for three problems we are trying to solve. One is how will the biased term b in equation (3) affect the transaction frequency, transaction cost, and profit. While maybe hedge funds may apply frequent trading strategies, many passive funds, alternative investment management funds, and individual investors may require fewer transaction frequencies. This is can be easily achieved by our robo-traders. Two is the performance between recurrent neural network versus the LSTM recurrent neural network. Last but not least, we will experiment not only maximizing downside deviation ratio and also feed in signals where the time series is consistently trending down and volatile signals to experiment protecting the down side of the investment. We'll also compare the performance for using Downside Deviation Ratio vs the Sharpe Ratio. These FX simulations demonstrate the ability of the Recurrent Reinforcement Learning algorithm to discover structure in a real world financial price series. Our goal is to find optimal solutions with few manual parameter tuning and can optimize performance regarding any trading period. 

To set up the experiment, it is implemented in Python with Pandas, Numpy, Sklearn, Keras and Tensorflow library. We take 2000 data points from US Dollar/British Pound price series, we used 1000 30 minutes interval price points from 1/6/17 to 2/3/17 as training set and again take 1000 30 minutes interval price points from 2/3/17 to 3/6/17 as test set. Looking back at equation (4), the first term of $sF_{t-1}r_{t}$ is the return from the investment decision from period t-1. For example, if s = 10 shares, the decision made by the trader is to buy half the maximum allowed and each share increased by $r_t$=2. This term would then be 10, which is the total return profit ignoring the transaction cost occurred during period t. The F term can be long, short or neutral or mathematically expressed as [-1,1] 

The chosen robo-trader we used in this experiment will be using LSTM along with Recurrent Reinforcement Learning with 55 percent dropout and continuous 200 data points of forecasting. We will further discuss in the second problem LSTM Recurrent Reinforcement Learning versus Recurrent Reinforcement Leanring.  We set the biased term one with b=1 and one with b=5. Comparing the biased term b=5 versus b=1, we can quickly notice that b=5 will results in less frequent transaction with less transaction cost resulting higher return during testing.

Figure 3. shows how Sharpe Ratio evolve through training, setting Epochs at 1000 times. Results of our various test runs that the optimized Sharpe Ratio increases as Epoch times increases. Intuitively, though it is easy to argue setting Epoch times extreme large to increase the final Sharpe Ratio may provide incremental performance, however our empirical results shows the performance or total profit in our test sets does not increase dramatically in the large epoch times case.% In fact, large epochs may cause over fitting and and reduce the performance versus initial weights. 

\begin{figure}[H]
%\DeclareGraphicsExtensions{.png,.pdf}
\centering
\graphicspath{{BR/}}
\includegraphics[width=3.3in]{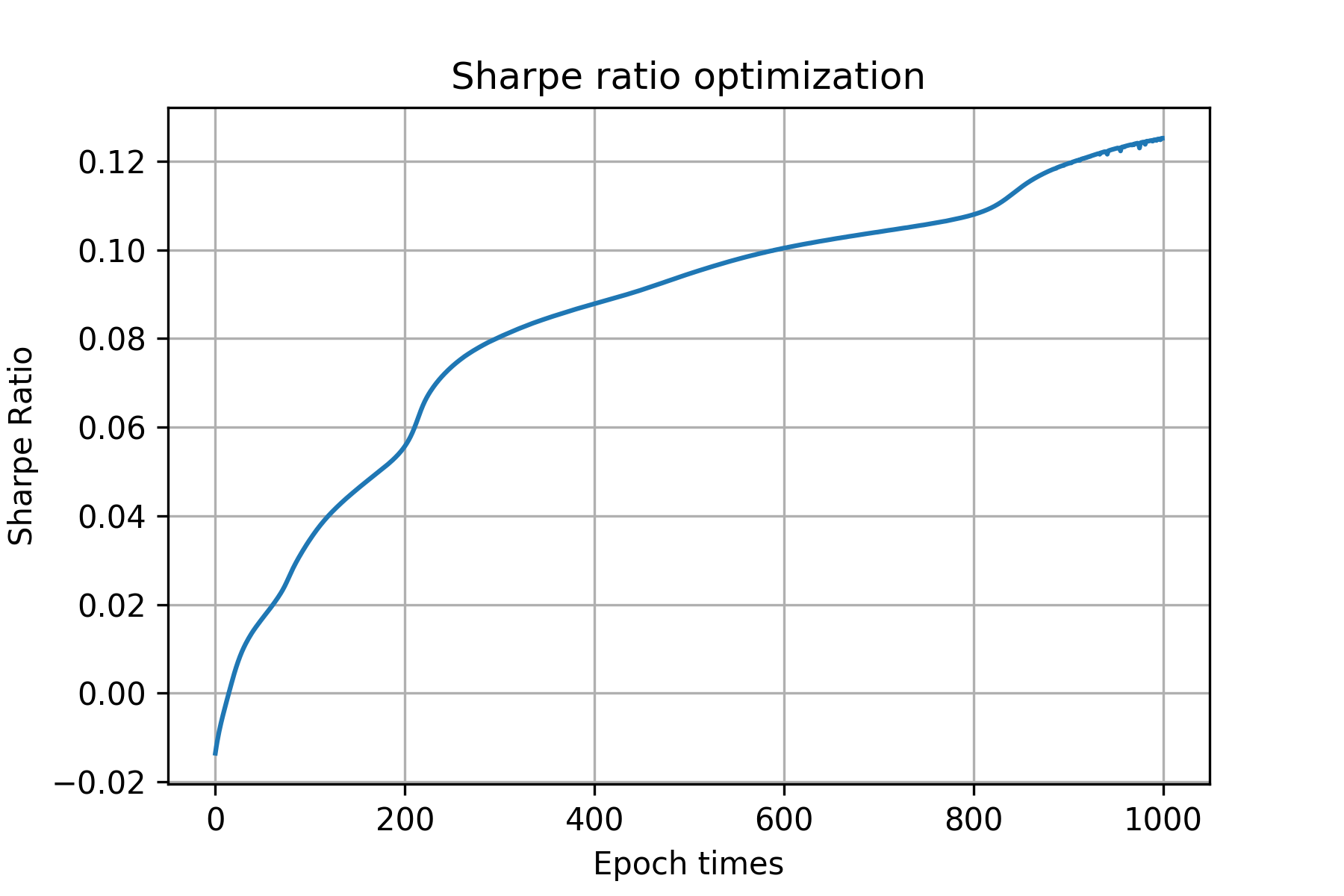}
% where an .eps filename suffix will be assumed under latex, 
% and a .pdf suffix will be assumed for pdflatex; or what has been declared
% via \DeclareGraphicsExtensions.
\captionsetup{justification=centering}
\caption{Sharpe Ratio through training}
%\label{fig_sim}
\end{figure}

The top panel of figure 4. shows the training set price of USDGBP from 1/6/17 to 2/3/17. The second panel shows trading signal produced by the trading system in training, and the third panel displays the profit and loss for the weights based on training. As we optimize the weights, the profit for training will gradually increase through out the training period.

\begin{figure}[h]
%\DeclareGraphicsExtensions{.png,.pdf}
\centering
\graphicspath{{BR/}}
\includegraphics[width=3.5in]{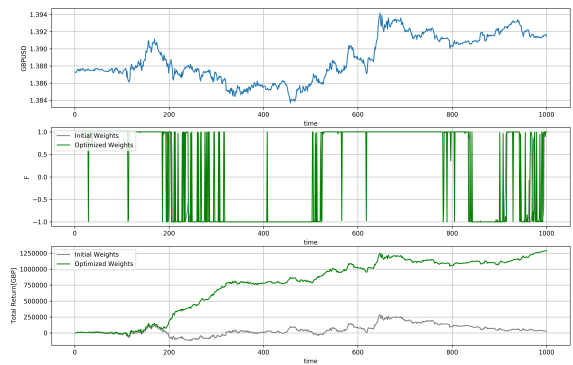}
% where an .eps filename suffix will be assumed under latex, 
% and a .pdf suffix will be assumed for pdflatex; or what has been declared
% via \DeclareGraphicsExtensions.
\captionsetup{justification=centering}
\caption{Training b = 1 - LSTM}
%\label{fig_sim}
\end{figure}

Similarly set up, the top panel of figure 5 shows the test set price of USDGBP from 2/3/17 to 3/6/17. The second panel shows trading signal produced by the trading system, and the bottom panel shows the profit and loss based on the initial weights vs trained weights. The performance for optimized weight with b=1 is actually worse than initial weights set up.

Figure 6. and 7. are the repeated training and testing for figure 4. and 5., respectively setting biased term to 5. As we can see in figure 5 and figure 7, with b set as 1 it's trading on an average of 6 hours per transaction while on with b set as 5, it's an average of 70 hours per transaction, which significantly reduce the transaction cost, therefore achieve better performance.

\begin{figure}[H]
%\DeclareGraphicsExtensions{.png,.pdf}
\centering
\graphicspath{{BR/}}
\includegraphics[width=3.5in]{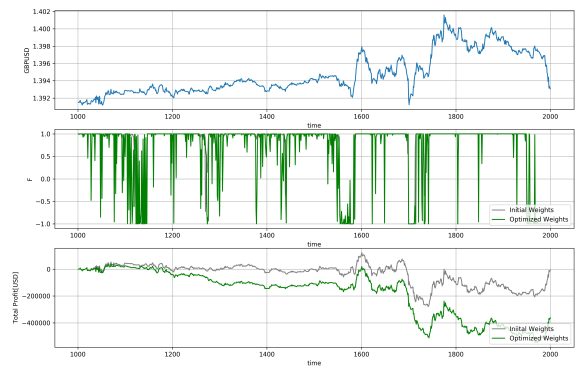}
% where an .eps filename suffix will be assumed under latex, 
% and a .pdf suffix will be assumed for pdflatex; or what has been declared
% via \DeclareGraphicsExtensions.
\captionsetup{justification=centering}
\caption{Prediction b = 1 - LSTM}
%\label{fig_sim}
\end{figure}

\begin{figure}[h]
%\DeclareGraphicsExtensions{.png,.pdf}
\centering
\graphicspath{{LSTM/}}
\includegraphics[width=3.5in]{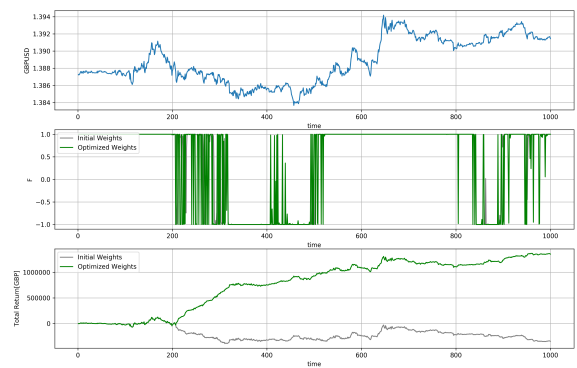}
% where an .eps filename suffix will be assumed under latex, 
% and a .pdf suffix will be assumed for pdflatex; or what has been declared
% via \DeclareGraphicsExtensions.
\captionsetup{justification=centering}
\caption{Training b = 5 - LSTM}
%\label{fig_sim}
\end{figure}

\begin{figure}[H]
%\DeclareGraphicsExtensions{.png,.pdf}
\centering
\graphicspath{{LSTM/}}
\includegraphics[width=3.5in]{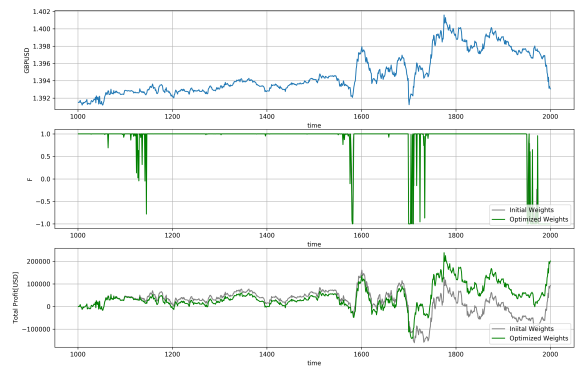}
% where an .eps filename suffix will be assumed under latex, 
% and a .pdf suffix will be assumed for pdflatex; or what has been declared
% via \DeclareGraphicsExtensions.
\captionsetup{justification=centering}
\caption{Prediction b = 5 - LSTM}
%\label{fig_sim}
\end{figure}

Second problem we would like to tackle is the performance between our LSTM Recurrent Reinforcement Learning trader vs Recurrent Reinforcement Learning trader. Comparing figure 8 versus figure 7, we can see our LSTM trader provides better performance with higher end total profit. 

\begin{figure}[H]
%\DeclareGraphicsExtensions{.png,.pdf}
\centering
\graphicspath{{RRL/}}
\includegraphics[width=3.3in]{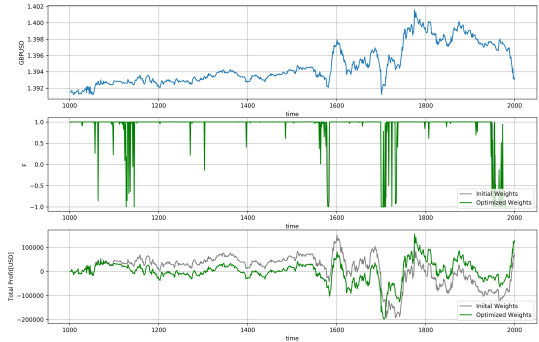}
% where an .eps filename suffix will be assumed under latex, 
% and a .pdf suffix will be assumed for pdflatex; or what has been declared
% via \DeclareGraphicsExtensions.
\captionsetup{justification=centering}
\caption{Prediction b = 5 RRL}
%\label{fig_sim}
\end{figure}

Lastly, the test for testing Downside Deviation Ratio. Figure 9. shows DDR training with Figure 10. showing the performance, which can be compared with Figure 7.

\begin{figure}[H]
%\DeclareGraphicsExtensions{.png,.pdf}
\centering
\graphicspath{{DDR/}}
\includegraphics[width=3.3in]{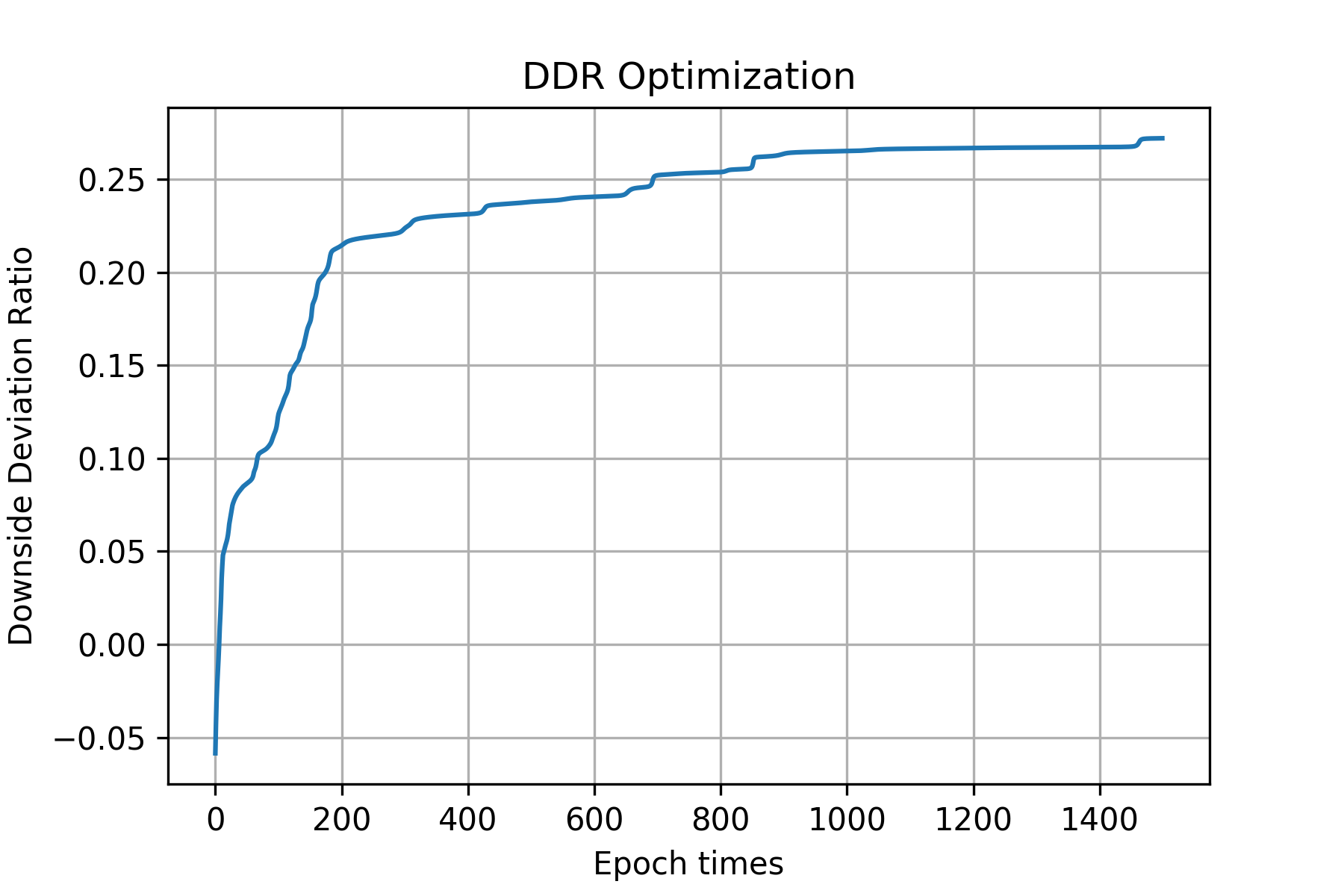}
% where an .eps filename suffix will be assumed under latex, 
% and a .pdf suffix will be assumed for pdflatex; or what has been declared
% via \DeclareGraphicsExtensions.
\captionsetup{justification=centering}
\caption{DDR through training}
%\label{fig_sim}
\end{figure}

\begin{figure}[H]
%\DeclareGraphicsExtensions{.png,.pdf}
\centering
\graphicspath{{DDR/}}
\includegraphics[width=3.5in]{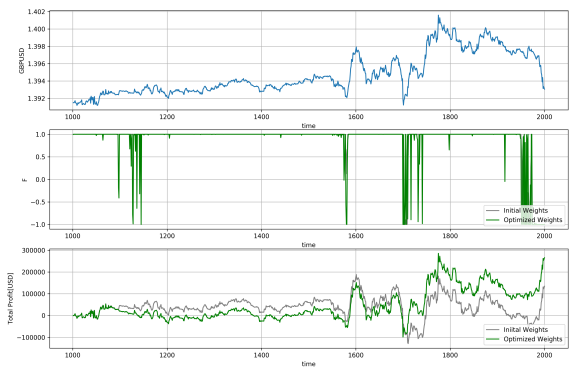}
% where an .eps filename suffix will be assumed under latex, 
% and a .pdf suffix will be assumed for pdflatex; or what has been declared
% via \DeclareGraphicsExtensions.
\captionsetup{justification=centering}
\caption{DDR b=5 -LSTM}
%\label{fig_sim}
\end{figure}

As we can observe, the robo-trader that were trained to maximize Downside Deviation Ratio has less loss or drawdowns then those maximized by Sharpe Ratio. We find that both robo-traders where successful in discovering profitable strategies, and that their learned behaviors exhibit risk-avoiding characteristics where benefits in a volatile market.

Though we aimed to pick a period in testing where exist low volatility and high volatility, in a real time trading system running 24 hours a day, it may suffer additional penalties when trying to trade during off-peak or low liquidity trading periods. An accurate test of this trading system would require live trading with a broker or directly through the interbank FX market with bid and ask price in order to verify real time transactable prices and profitability.

\section{Conclusion}
Like a human, our agents learn for themselves to achieve successful strategies that lead to the greatest long-term rewards. This paradigm of learning by trial-and-error, solely from rewards or punishments, is known as reinforcement learning (RL). Also like a human, our agents construct and learn their own knowledge directly from raw inputs, such as computer vision, without any hand or manual engineered features or domain heuristics. This is achieved by deep learning of neural networks. DeepMind had pioneered the combination of these approaches - deep reinforcement learning - to create the first artificial agents to achieve human-level performance in Alpha Go. There are many areas one can explore for deep learning and LSTM recurrent neural networks, areas such as prediction based models, classification based models and unsupervised learning. Our goal was to simply have AI to trade itself with downside protection and as few tweaks on the parameters as possible. With deep learning and recurrent networks we can explore feeding additional features such as volume, or interest rate, credit default swap spread and also provide multi-asset trading. 

 The empirical results presented controlled experiments using FX data with skewed returns distributions as a successful USDGBP trading agent. We've also shown that we can achieve downside protection with downside deviation ratio and feeding signals.This will help investors risk management when market sentiment is pessimistic. For passive investors, our results  suggest that long periodic strategy is feasible and can be used to combine with stock or other assets picking strategies. 
 
 The opportunities herein presented a powerful true robo-trading techniques where as few human intervene as possible. Though there are some parameters that would need careful selection future work included automating these type of dependencies with possible multi-asset strategies and real time trading mentioned throughout our discussions.

% if have a single appendix:
%\appendix[Proof of the Zonklar Equations]
% or
%\appendix  % for no appendix heading
% do not use \section anymore after \appendix, only \section*
% is possibly needed

% use appendices with more than one appendix
% then use \section to start each appendix
% you must declare a \section before using any
% \subsection or using \label (\appendices by itself
% starts a section numbered zero.)
%

%\appendices
%\section{Proof of the First Zonklar Equation}
%\blindtext

% use section* for acknowledgement
\section*{Acknowledgment}
The author would like to thank Professor Ching-Ho Leu from the Department of Statistics at National Cheng-Kung University, Tainan, Taiwan for his relentless mentorship and his helpful comments on this manuscript.

% Can use something like this to put references on a page
% by themselves when using endfloat and the captionsoff option.
\ifCLASSOPTIONcaptionsoff
  \newpage
\fi


\begin{thebibliography}{1}

%\bibitem{IEEEhowto:kopka}
%H.~Kopka and P.~W. Daly, \emph{A Guide to \LaTeX}, 3rd~ed.\hskip 1em plus
%  0.5em minus 0.4em\relax Harlow, England: Addison-Wesley, 1999.

\bibitem
 {1}R. Sutton, A. Barto, Reinforcement Learning: An Introduction, 2nd ed. Cambridge, Massachusetts, The MIT Press, 2012
\bibitem
 {2}R. Sutton, A. Barto, R. Williams, Reinforcement Learning is Direct Adaptive Optimal Control, IEEE Control Systems, April, 1992
\bibitem
 {3}R. Bellman, Dynamic Programming, Princeton University Press, Princeton, NJ, 1957
\bibitem
 {4}R. Sutton, Learning to Predict By the Method of Temporal Differences, Machine Learing, vol.3, 1988
\bibitem
 {5}C.J. Watkins, P.Dayan, Technical note: Q-Learning, Machine Learning, vol.9, 1992
\bibitem
 {6}L. Kaelbling, M. Littman, A. Moore, Reinforcment Learning" A Survey, Journal of Artificial Intelligence 
 Research, vol.4, 1996
\bibitem 
 {7}J. Moody, M. Saffell, Learning to Trade via Direct Reinforcement, IEEE Transactions on Neural Networks, Vol.12, July, 2001
\bibitem
 {8}D. Lu, Portfolio Optimization Using Linear Matrix Inequilities, IIT Working Paper, 2005
\bibitem
 {9}Y. Deng, F. Bao, Y. Kong, Z. Ren, Q. Dai, Deep Direct Reinforcement Learning for Financial Signal Representation and Trading, IEEE Transactions on Neural Networks and Learning Systems, April, 2015
\bibitem
 {10}W. Zaremba, I. Sutskever, O. Vinyals, Recurrent Neural Network Regularization, ICLR, 2015
\bibitem
 {11}H. Beyer, The Theory of Evolution Strategies, Springer-Verlag, New York, USA, 2001
\bibitem
 {12}C.J. Price, I.D. Coope, D. Byatt, A Convergent Variant of the Nelder-Maead Algorithm, J.Optim. Theory Appl. 113, No.1
\bibitem
 {13}J. Moody, L. Wu, Y. Liao, M. Saffell, Performance Functions and Reinforcement Learning for Trading Systems and Portfolios, Journal of Forecasting, vol. 17, 1998
\bibitem
 {14}W. Sharpe, The Sharpe Ratio, Journal of Portfolio Management, vol.21, 1994
\bibitem
 {15}C. Gold, FX Trading via Recurrent Reinforcement Learning, Computation and Neural Systems, 2003
\bibitem
 {16}P.J. Werbos, Back-Propagation Through Time: What it Does and How to Do it, IEEE Proceedings, vol.78, Oct, 1990
\bibitem
 {17}Y. Kao, B. Van Roy, Directed Principal Component Analysis, Operations Research, July, 2014
\bibitem
 {18}S. Hochreiter, J. Schmidhuber, Long Short-Term Memory, Neural Computation, 1997
\bibitem
 {19}Y. Bengio, P. Simard, P. Frasconi, Learning Long-Term Dependencies with Gradient Descent is Difficult, IEEE Transactions On Neural Networks, March, 1994
\bibitem
 {20}I. Goodfellow, Y. Bengio, A. Courville, Deep Learning, Book Draft for The MIT Press, 2015
\bibitem
 {21}F. Gers, N. Schraudolph, J. Schmidhuber, Learning to forget: Continual prediction with LSTM, Neural Computation, 2000
\bibitem
 {22}I. Danihelka, G. Wayne, B. Uria, Nal Kalchbrenner, Alex Graves, Associative Long Short-Term Memory, arXiv, 2015
\bibitem
 {23}N. Srivastava, G. Hinton, A. Krizhevsky, I. Sutskever, R. Salakhutdinov, Dropout: A Simple Way to Prevent Neural Networks from Overfitting, Journal of Machine Learning Research, 15,  2014
\bibitem
 {24}W. Zaremba, I. Sutskever, O. Vinyals, Recurrent Neural Network Regularization, arXiv, Feb, 2015
\bibitem
 {25} J.A. Nelder, R. Mead, A Simplex Method for Function Minimization, Computer Journal, Vol. 7, Issue 4, 1965
\bibitem
 {26}J. Kennedy, R. Eberhart, Swarm Intelligence. Morgan Kaufmann, San Francisco, CA, 2001
\bibitem
 {27}R. Y. Rubinstein, D. P. Kroese, The Cross-Entropy Method: A Uniﬁed Approach to Combinatorial Optimization, Monte-Carlo Simulation and Machine Learning (Information Science and Statistics), Springer, 2004
\bibitem
 {28}N. Hansen, A. Ostermeier, Completely Derandomized Self-Adaptation in Evolution Strategies, In Evolutionary Computation, 9, 2001
\bibitem
 {29}J. Schmidhuber, D. Wiestra, M. Gagliolo, F. Gomoze, Training Recurrent Networks by Evolino, Neural Computation, 19, 2007
\bibitem
 {30}H.P. Schwefel: Evolution and Optimum Seeking: New York: Wiley \& Sons 1995.
\bibitem
 {31}D. Wierstra, T. Schaul, J. Peters,J. Schmidhuber, Natural evolution strategies,  In Proceedings of the Congress on Evolutionary Computation,  IEEE Press, 2008.
\bibitem
 {32}S. Boyd, L. El Ghaoui, E. Feron, V. Balakrishnan, Linear Matrix Inequilities in System and Control Theory, Society for Industrial and Applied Mathematics, 1994

\end{thebibliography}
\end{document}